\documentclass[aps,floats]{revtex4}
\usepackage{amsmath,amssymb}
\usepackage{graphicx,epsfig}
\usepackage[greek,english]{babel}

\begin{document}
\bibliographystyle {plain}

\def\oppropto{\mathop{\propto}} 
\def\opsimeq{\mathop{\simeq}}
\def\opoverderline{\mathop{\overline}}
\def\operarrow{\mathop{\longrightarrow}}
\def\opsim{\mathop{\sim}}

\def\fig#1#2{\includegraphics[height=#1]{#2}}
\def\figx#1#2{\includegraphics[width=#1]{#2}}


\title{ Many-Body Localization : construction of the emergent local conserved operators \\  via block real-space renormalization} 


\author{ C\'ecile Monthus }
 \affiliation{Institut de Physique Th\'{e}orique, 
Universit\'e Paris Saclay, CNRS, CEA,
91191 Gif-sur-Yvette, France}

\begin{abstract}
A Fully Many-Body Localized (FMBL) quantum disordered system is characterized by the emergence of an extensive number of local conserved operators that prevents the relaxation towards thermal equilibrium. These local conserved operators can be seen as the building blocks of the whole set of eigenstates. In this paper, we propose to construct them explicitly via some block real-space renormalization. The principle is that each RG step diagonalizes the smallest remaining blocks and produces a conserved operator for each block. The final output for a chain of $N$ spins is a hierarchical organization of the $N$ conserved operators with $\left(\frac{\ln N}{\ln 2}\right)$ layers. The system-size nature of the conserved operators of the top layers is necessary to describe the possible long-ranged order of the excited eigenstates and the possible critical points between different FMBL phases. We discuss the similarities and the differences with the Strong Disorder RSRG-X method that generates the whole set of the $2^N$ eigenstates via a binary tree of $N$ layers. The approach is applied to the Long-Ranged Quantum Spin-Glass Ising model, where the constructed excited eigenstates are found to be exactly like ground states in another disorder realization, so that they can be either in the paramagnetic phase, in the spin-glass phase or critical.

\end{abstract}

\maketitle

\section{ Introduction} 

The field of Many-Body Localization (MBL) has attracted a lot of interest recently
 (see the recent reviews \cite{revue_huse,revue_altman} and references therein).
The general goal is to understand the unitary dynamics of isolated 
random interacting quantum systems 
and to determine how they can avoid thermalization and remain non-ergodic.

One of the most important characterization of the Many-Body localized phase
is that excited eigenstates display an area-law entanglement \cite{bauer}
instead of the volume-law entanglement of thermalized eigenstates. 
This property has been used numerically to identify the MBL phase
 \cite{kjall,alet} and to show the consistency with other criteria
of MBL \cite{alet}.
The fact that excited eigenstates in the
 MBL phase are similar to the ground-states
from the point of view of entanglement suggests 
that various approaches that have been developed for ground states in the past
can be actually adapted to study the MBL phase. 
One first example is the efficient representation
via  Density-Matrix-RG or Matrix Product States \cite{pekker1,pekker2,friesdorf,sondhi}
 and Tensor Networks \cite{tensor}.
Another example is the Strong Disorder Real-Space RG approach  
(see \cite{strong_review,refael_review} for reviews)
developed by Ma-Dasgupta-Hu \cite{ma_dasgupta}
 and Daniel Fisher \cite{fisher_AF,fisher}
to construct the ground states of random quantum spin chains,
with its extension in higher dimensions $d=2,3,4$ \cite{fisherreview,motrunich,lin,karevski,lin07,yu,kovacsstrip,kovacs2d,kovacs3d,kovacsentropy,kovacsreview}.
This approach has been extended into the Strong Disorder 
RG procedure for the unitary dynamics \cite{vosk_dyn1,vosk_dyn2},
and into the RSRG-X procedure
in order to construct the whole set of excited eigenstates 
 \cite{rsrgx,rsrgx_moore,vasseur_rsrgx,yang_rsrgx,rsrgx_bifurcation}.
 It should be stressed that these two Strong Disorder RG procedures
based on the spin variables are limited to the MBL phase,
whereas the current RG descriptions of the MBL transition towards delocalization 
 are based on RG rules for the entanglement \cite {vosk_rgentanglement}
 or for the resonances \cite{vasseur_resonant}.

Another recent essential idea in the field 
is the claim that a fully many-body localized system (FMBL)
can be characterized
 by an extensive number of emergent {\it localized } conserved operators
\cite{emergent_swingle,emergent_serbyn,emergent_huse,emergent_ent,imbrie,serbyn_quench,emergent_vidal,emergent_ros,emergent_rademaker}.
More precisely for a random quantum chain containing
 $N$ spins $\sigma_i$ described
 by Pauli matrices, it should be possible to introduce
 $N$ {\it localized } pseudo-spin operators $\tau_i^z$
that commute with each other and with the Hamiltonian.
In terms of these localized pseudo-spins $\tau_i$, 
the Hamiltonian can be rewritten as 
 \begin{eqnarray}
H^{FMBL} &&  = \sum_{k=0}^N \sum_{1 \leq i_1<i_2<..<i_k \leq N} 
J^{(k)}_{i_1,i_2,..,i_k} \prod_{q=1}^k \tau^z_{i_q} 
\label{hmbl}
\end{eqnarray}
containing $2^N$ independent couplings $J^{k}_{i_1,..,i_k}$. It should be stressed that the diagonalization
 of any Hamiltonian with $2^N$ energy levels can be rewritten as Eq. \ref{hmbl} since the $2^N$ couplings are sufficient to reproduce the $2^N$ energies.
So the non-trivial statement in the FMBL Hamiltonian of Eq. \ref{hmbl}
is that the pseudo-spins $\tau^z_i $ are obtained from the real spins $\sigma_i^z$
by a {\it quasi-local } unitary transformation, and that the couplings  
$J^{k}_{i_1,..,i_k}$ decay exponentially with the distance with a sufficient rate (see \cite{emergent_swingle,emergent_serbyn,emergent_huse,emergent_ent,imbrie,serbyn_quench,emergent_vidal,emergent_ros,emergent_rademaker} and the reviews \cite{revue_huse,revue_altman} for more details). 
On the contrary, in the delocalized phase, the pseudo-spins $\tau_i^z$ that
diagonalize the Hamiltonian are delocalized over the whole chain.

Since the essential property of these conserved pseudo-spins $\tau_i^z$ 
in the FMBL phase is their {\it locality}, 
it seems natural to try to generate them locally and homogeneously in space.
In the present paper, we thus propose to construct them
via some block Real-Space Renormalization procedures
that generalize previous procedures introduced for ground states.
The principle is that at each RG step, instead of projecting always onto
the lowest-energy subspace, one keeps the projection onto the lowest
energy subspace and the projection onto the highest energy subspace,
so that the choice between the two subspaces corresponds to a conserved pseudo spin.
In spirit, the idea is thus very close to
the RSRG-X method \cite{rsrgx,rsrgx_moore,vasseur_rsrgx,yang_rsrgx,rsrgx_bifurcation} where each decimation produces a bifurcation in the spectrum
and identifies a conserved pseudo-spin. 
In practice, the output is however somewhat different. 
For instance for a chain of $N$ spins with $2^N$ eigenstates :

(i)  the RSRG-X method produces a binary branching tree with $N$ layers, corresponding to the $N$ successive decimations,
so that at the end of the procedure,
 the $2^N$ leaves of the tree are the eigenstates.

(ii) the present block Real-Space approach
produces a hierarchical tree with $\left(\frac{\ln N}{\ln 2}\right)$ layers
for the conserved operators themselves :
the first RG step produces in parallel
 $\frac{N}{2} $ conserved operators
 associated to the blocks of $L=2$ sites,
the second RG step produces $\frac{N}{4} $ conserved operators
 associated to the blocks of $L=4$ sites, and so on, i.e.
the $k$ RG step produces $\frac{N}{2^k} $ conserved operators associated to the blocks of $L=2^k$ sites. So the first RG steps produce an extensive number of very local
 conserved operators as expected. 
However the few conserved operators associated 
to the last RG steps are system-size and not 'local' anymore.
 But their system-size nature is necessary to describe
the possible long-ranged order of the excited eigenstates 
and the possible critical points between different FMBL phases \cite{rsrgx,protected}.

The paper is organized as follows.
We consider the random quantum Ising chain 
with random transverse fields and with random couplings
that are either short-ranged in section \ref{sec_chain},
or long-ranged in section \ref{sec_dyson}.
Our conclusions are summarized in section \ref{sec_conclusion}.

\section{ Random quantum Ising chain }

\label{sec_chain}

The quantum Ising chain
\begin{eqnarray}
H=- \sum_{i}  h_i \sigma_{i}^x- \sum_{i}  J_{i,i+1} \sigma_{i}^z \sigma_{i+1}^z 
\label{h1d}
\end{eqnarray}
with random transverse fields $h_i$ 
and random nearest neighbor couplings $J_{i,i+1}$
is 'trivially' in the Fully Many-Body Localized phase, as a consequence of the Jordan-Wigner mapping onto free fermions. 
Nevertheless, we feel that it is useful as a toy model to see how the present approach works in the simplest possible case.
In this section,  we thus describe how the Fernandez-Pacheco
 self-dual block-RG 
 used to construct first the ground state of the pure chain \cite{pacheco,igloiSD}
and then the ground state of the random chain \cite{nishiRandom,us_pacheco,us_renyi,us_watermelon} can be extended to construct the whole set of eigenstates.

\subsection{ Diagonalization of the intra-block Hamiltonian }

For each block of two spins $(2i-1,2i)$, the intra-block Hamiltonian chosen by Fernandez-Pacheco \cite{pacheco,igloiSD,nishiRandom,us_pacheco,us_renyi,us_watermelon}
\begin{eqnarray}
H^{intra}_{(2i-1,2i)} &&  = - h_{2i-1} \sigma_{2i-1}^x - J_{2i-1,2i}   \sigma_{2i-1}^z \sigma_{2i}^z
\label{h1box}
\end{eqnarray}
commutes with $\sigma_{2i}^z$. So for each eigenvalue $S_{2i}=\pm 1$, 
the diagonalization of the effective Hamiltonian for the remaining odd spin $\sigma_{2i-1}$
\begin{eqnarray}
h^{eff}_{2i-1}(S_{2i})&&  = - h_{2i-1} \sigma_{2i-1}^x- J_{2i-1,2i} S_{2i} \sigma_{2i-1}^z
\label{h1eff}
\end{eqnarray}
leads to the two eigenvalues (that are independent of the value $S_{2i}=\pm 1$ as it should by symmetry)
\begin{eqnarray}
\lambda^{\pm}_{2i-1}(S_{2i}) = \pm \sqrt{ h_{2i-1}^2 +J_{2i-1,2i}^2  }
\label{blocklambda}
\end{eqnarray}
with the following corresponding eigenvectors
\begin{eqnarray}
 \vert \lambda^{-}_{2i-1}(S_{2i}) > && =
 \sqrt{ \frac{1 +  \frac{J_{2i-1}S_{2i}}{\sqrt{ h_{2i-1}^2 +J_{2i-1,2i}^2  }} }{2} }
\vert S_{2i-1}=+1 , S_{2i}>
+  \sqrt{ \frac{1 -  \frac{J_{2i-1}S_{2i}}{\sqrt{ h_{2i-1}^2 +J_{2i-1,2i}^2  }} }{2} } \vert S_{2i-1}=-1 , S_{2i}> 
\nonumber \\
\vert \lambda^{+}_{2i-1}(S_{2i}) > && = 
 - \sqrt{ \frac{1 -  \frac{J_{2i-1}S_{2i}}{\sqrt{ h_{2i-1}^2 +J_{2i-1,2i}^2  }} }{2} }
\vert S_{2i-1}=+1 , S_{2i}>
+  \sqrt{ \frac{1 +  \frac{J_{2i-1}S_{2i}}{\sqrt{ h_{2i-1}^2 +J_{2i-1,2i}^2  }} }{2} } \vert S_{2i-1}=-1 , S_{2i}> 
\label{blocklambda1eigen}
\end{eqnarray}

To construct the ground state \cite{pacheco,igloiSD,nishiRandom,us_pacheco,us_renyi,us_watermelon}, one chooses to keep only the two degenerate lowest states
$ \vert \lambda^{-}(S_{2i}=\pm) > $.
Here, as in the RSRG-X method \cite{rsrgx,rsrgx_moore,vasseur_rsrgx,yang_rsrgx,rsrgx_bifurcation}, we wish to keep also the two degenerate highest levels $ \vert \lambda^{+}(S_{2i}=\pm) > $.
It is thus convenient to introduce a pseudo-spin $\tau_{R(2i)}$ for the choice between these two possible energy levels
and a spin $\sigma_{R(2i)}$ to distinguish the two degenerate states within each energy level.
More precisely, the four eigenstates are relabeled as follows
\begin{eqnarray}
\vert \lambda^{-}_{2i-1}(+) > && \equiv \vert \tau_{R(2i)}^z = +, \sigma^z_{R(2i)}=+>  
\nonumber \\
\vert \lambda^{-}_{2i-1}(-) > && \equiv \vert \tau_{R(2i)}^z = +, \sigma^z_{R(2i)}=->  
\nonumber \\
\vert \lambda^{+}_{2i-1}(-) > && \equiv \vert \tau_{R(2i)}^z = -, \sigma^z_{R(2i)}=->  
\nonumber \\
\vert \lambda^{+}_{2i-1}(+) > && \equiv \vert \tau_{R(2i)}^z = -, \sigma^z_{R(2i)}=+>  
\label{relabel1d}
\end{eqnarray}
so that the intra-block Hamiltonian of Eq. \ref{h1box} depends only on the pseudo spin $\tau_{R(2i)} $ 
\begin{eqnarray}
 H_{(2i-1,2i)} && =- \Omega_{R(2i)}  \tau_{R(2i)}^z  
\label{projH}
\end{eqnarray}
with the field
\begin{eqnarray}
 \Omega_{R(2i)} &&  =  \sqrt{ h_{2i-1}^2 +J_{2i-1,2i}^2  }
\label{rgomega}
\end{eqnarray}

Since this procedure is applied independently to all blocks  $(2i-1,2i)$,
the total intra-block Hamiltonian reads in this new basis
\begin{eqnarray}
H^{intra} && =  \sum_{i}  H^{intra}_{(2i-1,2i)} 
 =  -  \sum_{i}  \Omega_{R(2i)}  \tau_{R(2i)}^z  
\label{hintrafinal}
\end{eqnarray}
For a chain of $N$ spins, this intra-block Hamiltonian
 depends only on the $\frac{N}{2} $ pseudo-spins
$ \tau_{R(2i)}^z $, whose flippings are associated to the 'large' fields
$\Omega_{R(2i)}$, but is completely independent of the 
$\frac{N}{2}$ spins $\sigma_{R(2i)} $, whose flippings have 'no cost' yet
when one takes into account only $ H^{intra}$.
In the following, we thus make the approximation
 that the pseudo-spins $ \tau_{R(2i)}^z  $ for $i=1,2,..,\frac{N}{2}$
are the first $\frac{N}{2}$ emergent local conserved operators.
This approximation is possible 
only when the initial transverse
 fields $h_i$ and/or the initial couplings $J_{i,i+1}$
are distributed with continuous probability distributions, so that there isn't any exact degeneracy between the fields $\Omega_{R(2i)} $ of Eq. \ref{rgomega}. (For instance for the pure chain where all fields $\Omega_{R(2i)} $
coincide, it is clear that this approximation would be meaningless).

\subsection{ Renormalization of the extra-block Hamiltonian }

The extra-block Hamiltonian defined in terms of the initial spins
\begin{eqnarray}
H^{extra} && = H-H^{intra} = - \sum_{i}    h_{2i} \sigma_{2i}^x - \sum_{i}  J_{2i-2} \sigma_{2i-2}^z  \sigma_{2i-1 }^z   
\label{hextra1d}
\end{eqnarray}
has to be rewritten
 in terms of the new spin variables $(\tau_{R(2i)},\sigma_{R(2i)})$
 introduced in Eq. \ref{relabel1d}.
Within the approximation that the pseudo-spins $ \tau_{R(2i)}^z  $
are emergent local conserved operators, one neglects all terms involving the flip operators
$ (\tau_{R(2i)}^x , \tau_{R(2i)}^y)  $, and one obtains
the following effective Hamiltonian for the remaining spins $\sigma^z_{R(2i)} $
\begin{eqnarray}
 H^{extra}_{eff} && =   - \sum_{i}   h_{R(2i)}    \sigma^x_{R(2i)} 
- \sum_{i}  J_{R(2i-2),R(2i)}(\tau^z_{R(2i)} )
  \sigma^z_{R(2i-2)}\sigma^z_{R(2i)}
\label{hextrafinal}
\end{eqnarray}
where the renormalized transverse field
\begin{eqnarray}
  h_{R(2i)} =  h_{2i} \frac{ h_{2i-1} }{\sqrt{ h_{2i-1}^2 +J_{2i-1,2i}^2  }}
\label{hr}
\end{eqnarray}
has decreased $h_{R(2i)} \leq h_{2i} $
and where the renormalized coupling
\begin{eqnarray}
 J_{R(2i-2),R(2i)}(\tau^z_{R(2i)} )  =\tau^z_{R(2i)} J_{2i-2,2i-1}\frac{ J_{2i-1,2i} }{\sqrt{ h_{2i-1}^2 +J_{2i-1,2i}^2  }}
\label{jr}
\end{eqnarray}
has also decreased in absolute value
$\vert  J_{R(2i-2),R(2i)}  \vert \leq \vert J_{2i-2,2i-1} \vert$.
The renormalized fields $h_{R(2i)}$
and couplings $ J_{R(2i-2),R(2i)}$ are thus typically smaller than the
initial fields and couplings, and thus also typically smaller than
the fields $\Omega_{R(2i)} $ of Eq. \ref{rgomega} associated to the conserved
pseudo-spins $\tau_{R(2i)}^{z}$.

Eqs \ref{hr} and \ref{jr} are the same as in the RG rules for the ground state \cite{nishiRandom,us_pacheco,us_renyi}
except for the presence of the pseudo spin $ \tau^z_{R(2i)} $ that can change the sign of the effective coupling between the spins $\sigma^z_{R(2i-2)}$ and $\sigma^z_{R(2i)}$.

\subsection{ Comparison with the RSRG-X method  }

The differences with the RSRG-X method are the following :

(i) here one makes an a-priori arbitrary choice with the couplings 
$(h_{2i-1},J_{2i-1,2i})$ concerning the odd spins, while the RSRG-X method
 selects the biggest coupling of the whole chain at each step.

(ii) here the renormalized field $\Omega_{R(2i)}=\sqrt{ h_{2i-1}^2 +J_{2i-1,2i}^2  } $
that appear also in the eigenstates of Eq. \ref{blocklambda1eigen}, and in
the RG rules of Eqs \ref{hr} and \ref{jr}, take into account
 any ratio $h_{2i-1}/J_{2i-1,2i}$ between the transverse field $h_{2i-1}$
and the coupling $J_{2i-1,2i}$,
  whereas in the RSRG-X method it is approximated by the maximal coupling,
namely either $h_{2i-1} $ or $J_{2i-1,2i} $.

As for the RSRG method for the gound state \cite{fisher_AF,fisher},
the RSRG-X method \cite{rsrgx}
 is expected to become asymptotically exact at large RG steps
when the RG flow is towards an Infinite Disorder Fixed Point, 
which happens at the critical point between two many-body-localized phase  \cite{rsrgx}.
However for the present goal to construct the extensive local conserved operators
 of the MBL phase, it is clear that the majority of the pseudo-spins are produced
by the first RG steps on small scales, and not by the asymptotic RG flow at large scales.
In a previous work \cite{us_renyi} concerning the Shannon and Renyi entropies
 of the ground state, that are also dominated by the first RG steps
on small scales, 
it was found that the block RG approach 
gives a better approximation of the multifractal dimensions than the RSRG.
So here also we expect that the better RG rules of (ii) overcompensate
 the drawback of the arbitrary choice of (i) and that the block RG will
produce a better approximation for the whole set of local conserved operators.

\subsection{ Iteration of the renormalization process  }

\label{iteration}

In summary, at the end of the first RG step, the spectrum of the $2^N$ levels of the chain containing $N$ spins
has been decomposed into $2^{\frac{N}{2}}$ groups of $2^{\frac{N}{2}} $ levels each.
A given group is labeled by the values of the $\frac{N}{2}$ pseudo-spins
$ \tau_{R(2i)}^z =\pm 1 $ and has for reference energy the contribution of $H^{intra}$ Eq. \ref{hintrafinal}
\begin{eqnarray}
E_{R^1} ( \{ \tau_{R(2i)}^z \}  ) && =  -  \sum_{i=1}^{2^{n-1}}  \Omega_{R(2i)}  \tau_{R(2i)}^z   
\label{erefr1}
\end{eqnarray}
With respect to this reference energy, the $2^{\frac{N}{2}} $ levels are those of the quantum Ising chain 
described by the effective Hamiltonian $H^{extra}_{eff}$ of Eq. \ref{hextrafinal}
 for the $\frac{N}{2}$ spins $\sigma^z_{R(2i)} $
with the smaller renormalized fields $h_{R(2i)} \leq h_{2i}$ and the 
 smaller renormalized couplings $\vert  J_{R(2i-2),R(2i)}  \vert \leq \vert J_{2i-2,2i-1} \vert$.

The RG procedure may be iterated as follows.
Each group of $2^{\frac{N}{2}} $ levels will be subdivided into 
$2^{\frac{N}{4}}$ sub-groups of $2^{\frac{N}{4}} $ levels each.
Each subgroup will be labeled by the $\frac{N}{2}$ pseudo-spins $ \tau_{R(2i)}^z =\pm 1 $
of the first generation that define the group and by the subsequent choice of the
$\frac{N}{4}$ pseudo-spins $ \tau_{R^2(4i)}^z =\pm 1 $ of the second RG step,
so that the reference energy of this subgroup is now
\begin{eqnarray}
E_{R^2} ( \{ \tau_{R(2i)}^z \};\{ \tau_{R^2(4i)}^z \} ) &&
 =  - \sum_{i=1}^{2^{n-1}}   \Omega_{R(2i)}  \tau_{R(2i)}^z  
 - \sum_{i=1}^{2^{n-2}} \Omega_{R^2(4i)}  \tau_{R^2(4i)}^z
\label{erefr2}
\end{eqnarray}
where the fields of the second generation
\begin{eqnarray}
 \Omega_{R^2(4i)} &&  =  \sqrt{ h_{R(2i-2)}^2 +J_{R(2i-2),R(2i)}^2  }
\label{rgomega2}
\end{eqnarray}
are actually independent of the multiplicative sign $\tau^z_{R(2i)} $ appearing in the renormalized coupling in Eq. \ref{jr}.
However the second generation pseudo-spin $\tau_{R^2(4i)}^z  $ itself depends on
the two pseudo-spins values $\tau_{R(4i-2)}^z=\pm 1$ and $\tau_{R(4i)}^z=\pm 1$ 
of the first generation. Of course, if one insists on reproducing the form of Eq. \ref{hmbl},
it is always possible to rewrite 
any function of these two pseudo-spins as the polynomial
\begin{eqnarray}
f (\tau_{R(4i-2)}^z,\tau_{R(4i)}^z)  
= f^{(0,0)} +f^{(1,0)} \tau_{R(4i-2)}^z+f^{(0,1)}\tau_{R(4i)}^z+f^{(1,1)}\tau_{R(4i-2)}^z \tau_{R(4i)}^z
\label{poly}
\end{eqnarray}
where the four coefficients are chosen to reproduce the only
four possibles values $f(\pm1,\pm1)$. 
But in the present framework,
it is simpler to stick to the pseudo spin $\tau_{R^2(4i)}^z $ 
as defined by the renormalization process in each sector, and to keep in mind that it
depends on the lower-generation pseudo-spins within its block.

With respect to the reference energy of Eq. \ref{erefr2}, the $2^{\frac{N}{4}} $ levels of this subgroup are those of the quantum Ising chain 
of $\frac{N}{4}$ spins $\sigma^z_{R^2(4i)} $ with the appropriate renormalized transverse field and coupling
obtained by the next iteration of the RG rules of Eq. \ref{hr}  and \ref{jr} 
\begin{eqnarray}
  h_{R^2(4i)}  &&=  h_{R(4i)} \frac{ h_{R(4i-2)} }{\sqrt{ h_{R(4i-2)}^2 +J_{R(4i-2),R(4i)}^2  }}
\nonumber \\
 J_{R^2(4i-4),R^2(4i)}  && =\tau^z_{R^2(4i)} J_{R(4i-4),R(4i-2)}\frac{ J_{R(4i-2),R(4i)} }{\sqrt{ h_{R(4i-2)}^2 +J_{R(4i-2),R(4i)}^2  }}
\label{rg2}
\end{eqnarray}

For a chain of $N=2^n$ spins, the procedure ends after the $n^{th}$ RG step :
the choice of $2^{n-1}$ pseudo spins $ \tau_{R(2i)}^z  $ of the first generation
has been followed by the choice of $2^{n-2}$ pseudo spins $ \tau_{R^2(4i)}^z  $ 
of the second generation, then by the choice of $2^{n-3}$ pseudo spins $ \tau_{R^2(8i)}^z  $ of the third generation, etc...
up to the choice of $2^0=1$ pseudo spin $ \tau_{R^n(2^n)}^z$ of the $n^{th}$ last generation;
 there remains a single spin $ \sigma_{R^n(2^n)} $ with its renormalized field $ h_{R^n(2^n)}$ so that the Hamiltonian is fully diagonalized and reads
\begin{eqnarray}
H_{R^n}  &&
 =  - \sum_{i=1}^{2^{n-1}} \Omega_{R(2i)}  \tau_{R(2i)}^z  
 - \sum_{i=1}^{2^{n-2}} \Omega_{R^2(4i)}  \tau_{R^2(4i)}^z 
- \sum_{i=1}^{2^{n-3}} \Omega_{R^3(8i)}  \tau_{R^3(8i)}^z 
\nonumber \\
&& ...
 - \Omega_{R^n(2^n)}  \tau_{R^n(2^n)}^z 
-  h_{R^n(2^n)}  \sigma_{R^n(2^n)}^z
\label{erefrn}
\end{eqnarray}
i.e. $\sigma_{R^n(2^n)}^z $ is the last $N^{th}$ conserved operator.

Note that Eq. \ref{erefrn} contains only $2^n=N$ parameters (namely 
$(2^n-1)$ parameters $\Omega_{R^k}$ of the various generations $1 \leq k \leq n$ and one parameter $ h_{R^n(2^n)} $) instead of the $2^N$ coefficients of Eq. \ref{hmbl} to parametrize the $2^N$ energies : this is a consequence of the free-fermion nature of the quantum Ising chain of Eq. \ref{h1d}.
For models that cannot be reduced to free-fermions, the diagonalized Hamiltonian
will be more complicated than Eq. \ref{erefrn}, as shown explicitely in section \ref{sec_dyson}.

\subsection{ Properties of the eigenstates  }

In terms of the ratio
\begin{eqnarray}
K_{i-1} \equiv \frac{J_{i-1,i}}{h_{i}} 
\label{ki}
\end{eqnarray}
the RG rules of Eq. \ref{hr} and \ref{jr} reduce to the simple multiplicative rule
\begin{eqnarray}
K_{R(2 i-2)} && \equiv \frac{J^R_{R(2i-2),R(2i)}}{h_{R(2i)}} =\tau^z_{R(2i)} K_{2i-2} K_{2i-1} 
\label{rgruleskr}
\end{eqnarray}
The conclusions for the amplitudes $\vert K \vert $ are thus exactly the same as for the ground state case corresponding to the choice $\tau^z_{R(2i)}=+1 $
discussed in detail in previous works \cite{nishiRandom,us_pacheco,us_renyi} :

(i) in the region $\overline{ \ln h_i} > \overline{ \ln J_{i,i+1}} $, the flow is attracted
towards $K \to 0$
and the eigenstates are paramagnetic.

(ii)  in the region $\overline{ \ln h_i} < \overline{ \ln J_{i,i+1}} $, 
the flow is attracted towards $\vert K \vert \to +\infty$
and the eigenstates display a long-ranged order adapted to the signs $\tau^z=\pm 1 $.
In particular in the middle of the spectrum, 
as a consequence of the random signs $\tau^z=\pm 1 $,
the eigenstates are in the spin-glass phase for any initial distribution of the couplings
(even if the initial model is ferromagnetic $J_{i,i+1}>0$).

(iii) at the phase transition $\overline{ \ln h_i} = \overline{ \ln J_{i,i+1}} $, 
the eigenstates are critical and described by
the Infinite Disorder Fixed point with the activated exponent $\psi=1/2$,
with the typical correlation length exponent $\nu_{typ}=1 $
and the finite-size correlation exponent $\nu_{FS}=2$,  in agreement with the Fisher Strong Disorder exact results \cite{fisher}.
 
(iv) the eigenstates have also exactly the same Shannon-R\'enyi entropies as the ground state studied in \cite{us_renyi}, since the signs $\tau_i^z$ completely disappear from the RG rules
for the Shannon-R\'enyi entropies.

The entanglement properties of the excited eigenstates 
are also exactly the same as for the ground state,
namely an area law (i.e. a constant here in $d=1$) outside criticality,
and a logarithmic violation at criticality, whose exact behavior has been 
 computed via the Strong Disorder RG approach \cite{refael,review_entang}.

\section{ Long-Ranged Dyson Quantum Ising model }

\label{sec_dyson}

Many-Body Localization has been studied for various long-ranged 
power-law interactions \cite{levitov,burin_eloc,pino,yao,hauke,haas,burin,burinxy,LRexp}.
Any long-ranged model has a Dyson hierarchical analog where
 real space renormalization procedures are easier to define and to study: since its introduction for
the classical ferromagnetic Ising model \cite{dyson,bleher,gallavotti,book,jona,baker,mcguire,Kim,Kim77,us_dysonferrodyn}, this framework
 has been used recently for many long-ranged disordered models,
either classical like random fields Ising models \cite{randomfield,us_aval,decelle}
and spin-glasses \cite{franz,castel_etal,castel_parisi,castel,angelini,guerra,barbieri}, or quantum like Anderson localization models \cite{bovier,molchanov,krit,kuttruf,fyodorov,EBetOG,fyodorovbis,us_dysonloc} and quantum spin models \cite{c_dysonquantumferro,c_dysonquantumsg}.
In this section, we consider the Dyson hierarchical version 
of the Long-Ranged Quantum Ising Chain
\begin{eqnarray}
H &&  = - \sum_{i} h_i \sigma_i^z - \sum_{i<j}  J_{i,j}   \sigma_{i}^x \sigma_{j}^x
\label{hlr}
\end{eqnarray}
where the properties of the long-ranged couplings $J_{i,j}$, namely their decay with the distance $(j-i)$ and their signs, have to be specified.

\subsection{ Dyson hierarchical version of the Long-Ranged Quantum Ising Chain }

The Dyson quantum Ising Hamiltonian for $N=2^n$ spins
is defined as a sum over the generations $k=0,1,..,n-1$
\begin{eqnarray}
H_{(1,2^n)} &&  = \sum_{k=0}^{n-1} H^{(k)}_{(1,2^n)} 
\label{recDyson}
\end{eqnarray}
The Hamiltonian of lowest generation $k=0$ contains the 
transverse fields $h_i$ and the lowest order couplings $J^{(0)}_{2i-1,2i}$
associated to the length $L_0=2^0=1$ 
\begin{eqnarray}
 H^{(k=0)}_{(1,2^n)} &&  = - \sum_{i=1}^{2^n} h_i \sigma_i^z
 - \sum_{i=1}^{2^{n-1}}  J^{(0)}_{2i-1,2i}   \sigma_{2i-1}^x \sigma_{2i}^x
\label{h0dyson}
\end{eqnarray}
The Hamiltonian of next generation $k=1$ contains 
 couplings $J^{(1)}$ associated to the length $L_1=2^1=2$
\begin{eqnarray}
 H^{(k=1)}_{(1,2^n)} &&  =
 - \sum_{i=1}^{2^{n-2}}  
\left[ J^{(1)} _{4i-3,4i-1}   \sigma_{4i-3}^x \sigma_{4i-1}^x
+  J^{(1)} _{4i-3,4i}   \sigma_{4i-3}^x \sigma_{4i}^x
+ J^{(1)} _{4i-2,4i-1}   \sigma_{4i-2}^x \sigma_{4i-1}^x
+  J^{(1)} _{4i-2,4i}   \sigma_{4i-2}^x \sigma_{4i}^x
 \right]
\label{h1dyson}
\end{eqnarray}
and so on up to the last generation $k=n-1$ 
 associated to the length $L_{n-1}=2^{n-1}=\frac{N}{2}$
that couples 
all pairs of spins between the two halves of the system
\begin{eqnarray}
 H^{(n-1)}_{(1,2^n)} = - \sum_{i=1}^{2^{n-1}} \sum_{j=2^{n-1}+1}^{2^n} 
 J^{(n-1)}_{i,j}  \sigma_i^x   \sigma_j^x 
\label{hlastdyson}
\end{eqnarray}

To mimic a power-law decay of the non-hierarchical model of Eq. \ref{hlr}
\begin{eqnarray}
J(r) \propto \frac{1}{r^{a}}
\label{jpower}
\end{eqnarray}
one chooses the same dependence for the scale of
the Dyson coupling $J^{(k)}$ of generation $k$
as a function of the length $L_k=2^k$
\begin{eqnarray}
J^{(k)} \propto \frac{1}{L_k^{a}} = 2^{-ka}
\label{jrk}
\end{eqnarray}
Within a given generation $k$, the couplings $J^{(k)}_{i,j}$
may be chosen uniform $J^{(k)}_{i,j}=J^{(k)}=2^{-ka}$ as in the 
Dyson quantum ferromagnetic Ising model
 with uniform or random transverse fields
discussed in \cite{c_dysonquantumferro} or may be taken as random variables
as in the Dyson quantum Spin-Glass discussed in \cite{c_dysonquantumsg}.
In the following, we extend the real-space RG procedure introduced to construct the ground
state of these models \cite{c_dysonquantumferro,c_dysonquantumsg}
in order to construct also the excited states.

\subsection{ Diagonalization of the lowest generation $k=0$ }

The elementary renormalization step 
concerns the block two-spin Hamiltonian of generation $k=0$ 
\begin{eqnarray}
H_{(2i-1,2i)} &&  = - h_{2i-1} \sigma_{2i-1}^z- h_{2i} \sigma_{2i}^z
- J^{(0)}_{2i-1,2i}   \sigma_{2i-1}^x \sigma_{2i}^x
\label{hbox}
\end{eqnarray}

The diagonalization within the symmetric sector
\begin{eqnarray}
H_{(2i-1,2i)} \vert ++> && = - (h_{2i-1}+h_{2i}) \vert ++ >-  J^{(0)}_{2i-1,2i} \vert -- >
\nonumber \\
H_{(2i-1,2i)} \vert --> && =-  J^{(0)}_{2i-1,2i} \vert ++ >  +(h_{2i-1}+h_{2i})  \vert -- >
\label{huvs}
\end{eqnarray}
yields the two eigenvalues  
\begin{eqnarray}
\lambda_{2i}^{S\pm} && = \pm \sqrt{ (J^{(0)}_{2i-1,2i})^2+(h_{2i-1}+h_{2i})^2 }
\label{lambdas}
\end{eqnarray}
corresponding to the eigenvectors
\begin{eqnarray}
\vert \lambda_{2i}^{S-} > && = \cos \theta_{2i}^S\vert ++>+\sin \theta_{2i}^S\vert -- >
\nonumber \\
\vert \lambda_{2i}^{S+} > && = -\sin \theta_{2i}^S\vert ++>+\cos \theta_{2i}^S\vert -- >
\label{vlambdas}
\end{eqnarray}
in terms of the angle $\theta_{2i}^S$ defined by
\begin{eqnarray}
\cos ( \theta_{2i}^S ) && = 
\sqrt{ \frac{1+ \frac{h_{2i-1}+h_{2i}}{ \sqrt{ (J^{(0)}_{2i-1,2i} )^2+(h_{2i-1}+h_{2i})^2 }}}{2}}
\nonumber \\
\sin ( \theta_{2i}^S ) && = {\rm sgn} (J^{(0)}_{2i-1,2i})
\sqrt{ \frac{1- \frac{h_{2i-1}+h_{2i}}{ \sqrt{ (J^{(0)}_{2i-1,2i} )^2+(h_{2i-1}+h_{2i})^2 }}}{2}}
\label{thetas}
\end{eqnarray}

Similarly the diagonalization within the antisymmetric sector
\begin{eqnarray}
H_{(2i-1,2i)} \vert +-> && = - (h_{2i-1}-h_{2i}) \vert +- >-  J^{(0)}_{2i-1,2i} \vert -+ >
\nonumber \\
H_{(2i-1,2i)} \vert -+> && =-  J^{(0)}_{2i-1,2i} \vert +- >  +(h_{2i-1}-h_{2i})  \vert -+ >
\label{huva}
\end{eqnarray}
leads to the two eigenvalues 
\begin{eqnarray}
\lambda_{2i}^{A\pm} && = \pm \sqrt{ (J^{(0)}_{2i-1,2i} )^2+(h_{2i-1}-h_{2i})^2 }
\label{lambdaa}
\end{eqnarray}
with the corresponding eigenvectors 
\begin{eqnarray}
\vert \lambda_{2i}^{A-} > && = \cos \theta_{2i}^A\vert +->+\sin \theta_{2i}^A\vert -+ >
\nonumber \\
\vert \lambda_{2i}^{A+} > && = -\sin \theta_{2i}^A\vert +->+\cos \theta_{2i}^A\vert -+ >
\label{vlambdaa}
\end{eqnarray}
in terms of the angle $\theta_A$ defined by
\begin{eqnarray}
\cos (  \theta_{2i}^A ) && =
\sqrt{ \frac{1+ \frac{h_{2i-1}-h_{2i}}{ \sqrt{ (J^{(0)}_{2i-1,2i} )^2+(h_{2i-1}-h_{2i})^2 }}
 }{2}}
\nonumber \\
\sin (  \theta_{2i}^A ) && = {\rm sgn} (J^{(0)}_{2i-1,2i})
\sqrt{ \frac{1- \frac{h_{2i-1}-h_{2i}}{ \sqrt{ (J^{(0)}_{2i-1,2i} )^2+(h_{2i-1}-h_{2i})^2 }}
 }{2}}
\label{thetaa}
\end{eqnarray}

To construct the ground states, one chooses to keep only the two negative energy levels
$ \lambda_{2i}^{S-}   $ and $ \lambda_{2i}^{A-}  $ \cite{c_dysonquantumferro,c_dysonquantumsg}.
Here we wish to keep also the two positive energy levels $ \lambda_{2i}^{S+}   $ and $ \lambda_{2i}^{A+}  $.
As in the previous section, it is thus convenient
to introduce a pseudo-spin $\tau_{R(2i)}^z=\pm$
for the choice between the subspace of positive or negative eigenvalues,
and another spin $\sigma_{R(2i)} $ for the subsequent choice between
the symmetric and the anti-symmetric sectors. More precisely,
we relabel the four eigenstates as follows
\begin{eqnarray}
\vert \lambda_{2i}^{S-} > && \equiv \vert \tau_{R(2i)}^z = +, \sigma^z_{R(2i)}=+>  
\nonumber \\ 
\vert \lambda_{2i}^{A-} > && \equiv \vert \tau_{R(2i)}^z=+, \sigma^z_{R(2i)}=->  
\nonumber \\ 
\vert \lambda_{2i}^{A+} > && \equiv \vert \tau_{R(2i)}^z=- ,\sigma^z_{R(2i)}=+>  
 \nonumber \\ 
\vert \lambda_{2i}^{S+} > && \equiv \vert \tau_{R(2i)}^z=-, \sigma^z_{R(2i)}=->   
\label{sigmaR2states}
\end{eqnarray}

The block two-spin Hamiltonian of Eq. \ref{hbox} then reads 
\begin{eqnarray}
 H_{(2i-1,2i)} && =
 - \Omega_{R(2i)}  \tau_{R(2i)}^z  -  h_{R(2i)} \sigma^z_{R(2i)}
\label{projHdyson}
\end{eqnarray}
where the renormalized field for $\tau_{R(2i)}^z$
\begin{eqnarray}
 \Omega_{R(2i)} &&  = \frac{\sqrt{ (J^{(0)}_{2i-1,2i})^2+(h_{2i-1}+h_{2i})^2 }+\sqrt{ (J^{(0)}_{2i-1,2i})^2+(h_{2i-1}-h_{2i})^2 }}{2} 
\label{rgB}
\end{eqnarray}
is bigger than the renormalized field for $\sigma_{R(2i)}^z$
\begin{eqnarray}
 h_{R(2i)} &&  = \frac{\sqrt{ (J^{(0)}_{2i-1,2i})^2+(h_{2i-1}+h_{2i})^2 }-\sqrt{ (J^{(0)}_{2i-1,2i})^2+(h_{2i-1}-h_{2i})^2 }}{2} 
\nonumber \\ 
&& = \frac{ 2 h_{2i-1} h_{2i} }
{\sqrt{ (J^{(0)}_{2i-1,2i})^2+(h_{2i-1}+h_{2i})^2 }
+ \sqrt{ (J^{(0)}_{2i-1,2i})^2+(h_{2i-1}-h_{2i})^2 }}
\label{rgh}
\end{eqnarray}
In the following, we thus make the approximation
 that the pseudo-spins $ \tau_{R(2i)}^z  $
are the first $\frac{N}{2}=2^{n-1}$ emergent local conserved operators.
As in the previous section, this approximation is possible 
only when the initial transverse fields $h_i$ and/or the initial couplings $J_i^{(0)}$
are distributed with continuous probability distributions, so that there isn't any exact degeneracy between the fields $\Omega_{R(2i)} $ of Eq. \ref{rgB}. 

\subsection{ Renormalization of the couplings of higher generations $k>0$ }

We need to compute the initial operators $\sigma^x_i$ in the new basis
of Eq. \ref{sigmaR2states}.
 Within the approximation that the pseudo-spins $ \tau_{R(2i)}^z  $
are emergent local conserved operators, 
one neglects all terms involving the flip operators
$ (\tau_{R(2i)}^x , \tau_{R(2i)}^y)  $, and one obtains
\begin{eqnarray}
\sigma_{2i-1}^x && \simeq c_{2i-1}  \tau_{R(2i)}^z \sigma_{R(2i)}^x
\nonumber \\
\sigma_{2i}^x && \simeq c_{2i} \sigma_{R(2i)}^x
\label{sxoddop}
\end{eqnarray}
in terms of the constants
\begin{eqnarray}
c_{2i-1} && = ( \cos \theta_{2i}^S \sin \theta_{2i}^A + \sin \theta_{2i}^S\cos \theta_{2i}^A) 
= {\rm sgn} (J^{(0)}_{2i-1,2i} )  \sqrt{ \frac{1+\frac{(J^{(0)}_{2i-1,2i})^2-h_{2i-1}^2+h_{2i}^2 }
{\sqrt{ (J^{(0)}_{2i-1,2i})^2+(h_{2i-1}+h_{2i})^2 }
\sqrt{ (J^{(0)}_{2i-1,2i})^2+(h_{2i-1}-h_{2i})^2 }}}{2}  }
\nonumber \\
c_{2i} && = ( \cos \theta_{2i}^S \cos \theta_{2i}^A - \sin \theta_{2i}^S\sin \theta_{2i}^A)
=  \sqrt{ \frac{1+\frac{(J^{(0)}_{2i-1,2i})^2+h_{2i-1}^2-h_{2i}^2 }
{\sqrt{ (J^{(0)}_{2i-1,2i})^2+(h_{2i-1}+h_{2i})^2 }
\sqrt{ (J^{(0)}_{2i-1,2i})^2+(h_{2i-1}-h_{2i})^2 }}}{2}  }
\label{cc}
\end{eqnarray}
As a consequence,
the renormalized coupling between the spins
$\sigma_{R(2i)} $ and $\sigma_{R(2j)} $
 is given by the following linear combination of the four initial couplings
of generation $k$ associated to the positions $(2i-1,2i)$ and $(2j-1,2j)$
\begin{eqnarray}
 J_{R(2i),R(2j)} && = J^{(k)}_{2i,2j} c_{2i} c_{2j}
+ \tau_{R(2i)}^z J^{(k)}_{2i-1,2j} c_{2i-1}  c_{2j}
+\tau_{R(2j)}^z J^{(k)}_{2i,2j-1} c_{2i} c_{2j-1}
+ \tau_{R(2i)}^z \tau_{R(2j)}^z J^{(k)}_{2i-1,2j-1} c_{2i-1}  c_{2j-1} 
\label{rgjbox}
\end{eqnarray}
So here in contrast with the short-ranged RG rule of Eq. \ref{jr},
the choice of the pseudo-spins values $(\tau_{R(2i)}^z=\pm 1;\tau_{R(2j)}^z=\pm1)$
will produce different amplitudes for the renormalized coupling in the different sectors.

\subsection{ Renormalization procedure for the non-hierarchical long-ranged model  }

The block renormalization procedure explained above can be applied to the 
non-hierarchical long-ranged model of Eq. \ref{hlr}. The only difference
is that in the RG rule of Eq. \ref{rgjbox}, 
the initial couplings have not exactly the same properties:
$J_{2i,2j} $ and $J_{2i-1,2j-1} $ are associated to the distance $(2j-2i)$,
whereas $J_{2i,2j-1} $ is associated to the sligthly smaller distance $(2j-2i-1)$
and $J_{2i-1,2j}$ is associated to the sligthly bigger distance $(2j-2i+1)$.
Whenever the long-range nature dominates over the short-range case,
one expect that the initial power-law model of Eq. \ref{hlr} and its Dyson
hierarchical analog have the same scaling properties at large scale.
However the RG flows for the Dyson analog are usually easier 
to study analytically and clearer to analyze numerically.

\subsection{ Iteration of the renormalization process  }

When the renormalization process is iterated as explained in section
\ref{iteration} with the appropriate renormalization rules of Eqs \ref{rgB}, \ref{rgh} and \ref{rgjbox},
it should be stressed that not only the pseudo-spins of higher generation depend on the lower pseudo-spins, 
but also the amplitudes of the renormalized fields and couplings.
More precisely, we may write the Hamiltonian
after the first RG step
\begin{eqnarray}
H^R_{(1,2^n)} && = - \sum_{i=1}^{2^{n-1}} \Omega_{R(2i)}  \tau_{R(2i)}^z 
 - \sum_{i=1}^{2^{n-1}} h_{R(2i)}  \sigma_{R(2i)}^z 
- \sum_{i <j} J_{R(2i),R(2j)}(\tau_{R(2i)}^z ,\tau_{R(2j)}^z ) \sigma_{R(2i)}^x \sigma_{R(2j)}^x 
\label{HamilR1}
\end{eqnarray}
after the second RG step
\begin{eqnarray}
&& H^{R^2}_{(1,2^n)}  =- \sum_{i=1}^{2^{n-1}} \Omega_{R(2i)}  \tau_{R(2i)}^z
  - \sum_{i=1}^{2^{n-2}} \Omega_{R^2(4i)}(\tau_{R(4i-2)}^z,\tau_{R(4i)}^z)  \tau_{R^2(4i)}^z
\label{HamilR2} \\
&&  - \sum_{i=1}^{2^{n-2}} h_{R^2(4i)}(\tau_{R(4i-2)}^z,\tau_{R(4i)}^z)  \sigma_{R(4i)}^z 
- \sum_{i <j} J_{R(4i),R(4j)}(\tau_{R(4i-2)}^z,\tau_{R(4i)}^z ,\tau_{R(4j-2)}^z,\tau_{R(4j)}^z,
 \tau_{R^2(4i)}^z ,  \tau_{R^2(4j)}^z ) \sigma_{R(4i)}^x \sigma_{R(4j)}^x 
\nonumber
\end{eqnarray}
and so on up the last $n^{th}$ RG step
\begin{eqnarray}
H^{R^n}_{(1,2^n)} && =- \sum_{i=1}^{2^{n-1}} \Omega_{R(2i)}  \tau_{R(2i)}^z
  - \sum_{i=1}^{2^{n-2}} \Omega_{R^2(4i)}(\tau_{R(4i-2)}^z,\tau_{R(4i)}^z)  \tau_{R^2(4i)}^z 
\nonumber \\
&&  - \sum_{i=1}^{2^{n-3}} \Omega_{R^3(8i)}(\tau_{R^2(8i-4)}^z,\tau_{R^2(8i)}^z;\tau_{R(4i-6)}^z,\tau_{R(8i-4)}^z, \tau_{R(8i-2)}^z,\tau_{R(8i)}^z)  \tau_{R^3(8i)}^z 
- ...
\nonumber \\
&& 
 - \Omega_{R^n(2^n)}( \{ \tau_{R^k}^z \} )  \tau_{R^n(2^n)}^z
-  h_{R^n(2^n)}( \{ \tau_{R^k}^z \}  )  \sigma_{R^n(2^n)}^z 
\label{HamilRfinal}
\end{eqnarray}

To make the link with Eq. \ref{hmbl}, one needs to expand the renormalized fields $\Omega_{R^p} $ and $h_{R^n(2^n)} $ in terms of the pseudo-spins of lower generations using the principle of Eq. \ref{poly} : 
each renormalized field of the second generation can be expanded into $2^2=4$ terms
\begin{eqnarray}
 \Omega_{R^2(4i)}(\tau_{R(4i-2)}^z,\tau_{R(4i)}^z)  
= \Omega_{R^2(4i)}^{(0,0)} +\Omega_{R^2(4i)}^{(1,0)} \tau_{R(4i-2)}^z+\Omega_{R^2(4i)}^{(0,1)}\tau_{R(4i)}^z+\Omega_{R^2(4i)}^{(1,1)}\tau_{R(4i-2)}^z \tau_{R(4i)}^z
\label{poly2}
\end{eqnarray}
each renormalized field $\Omega_{R^3(8i)}(\tau_{R^2(8i-4)}^z,\tau_{R^2(8i)}^z;\tau_{R(4i-6)}^z,\tau_{R(8i-4)}^z, \tau_{R(8i-2)}^z,\tau_{R(8i)}^z) $ of the third generation can be expanded into $2^6 $ terms
 and so on. This rewriting will thus generate all possible products containing an arbitrary number of pseudo-spins as in Eq. \ref{hmbl}.

As a consequence of the random sum structure of Eq. \ref{HamilRfinal}
over all blocks and all scales, 
one expects that at the middle of the spectrum near zero energy, two energy levels
that happen to be consecutive (i.e. that have an exponentially small energy difference
of order $(N^{\frac{1}{2}} 2^{-N})$ with respect to the number $N=2^n$ spins)
have completely different wave-functions
 labelled by completely different values of the pseudo-spins.

\subsection{ Properties of the eigenstates for the  Dyson quantum spin-glass model}

To be more specific, let us now focus on the case of the Dyson quantum spin-glass model, 
where 
the initial couplings $J^{(k)}_{i,j}$ of generation $k$ are random Gaussian variables
of zero mean and of variance
\begin{eqnarray}
\overline{ (J^{(k)}_{i,j})^2 }= 2^{-2 k \sigma} =  \frac{1}{L_k^{2\sigma}} 
\label{jndysonsg}
\end{eqnarray}
The parameter $\sigma$ governing the decay with the distance
has to be in the region $\sigma>1/2$ in order to have an extensive energy.
The initial transverse fields are taken uniform $h_i=h$  (but of course the renormalized transverse fields are
random as a consequence of the RG rule of Eq. \ref{rgh}).
For an excited state, the random choice of
the pseudo-spins $\tau^z_{R(2i)}=\pm$ in the RG rule
of Eq. \ref{rgjbox} simply amounts to randomly change the signs of the couplings that are
already of random sign. As a consequence, the RG flow is exactly the same as for the the ground state
 in another disorder realization.  
We can thus directly use the results concerning the RG flows for the ground state
studied in \cite{c_dysonquantumsg} as a function of the control parameter $h$ :  
the constructed excited eigenstates are similarly either in the paramagnetric phase, in the spin-glass phase or critical with a finite dynamical exponent $z(\sigma)$.

\section{ Conclusion }

\label{sec_conclusion}

In this paper, we have proposed to construct 
the extensive number of local conserved operators characterizing
a Fully Many-Body Localized quantum disordered system 
via some block real-space renormalization.
The general idea is that each RG step diagonalizes the smallest remaining blocks
 and produces a conserved pseudo-spin for each block:
for a chain of $N$ spins, the first RG step produces
 $\frac{N}{2} $ conserved operators associated to the blocks of $L=2$ sites, the second RG step produces $\frac{N}{4} $ conserved operators
 associated to the blocks of $L=4$ sites, and so on.
 The final output for a chain of $N$ spins is a hierarchical organization of the conserved operators with $\left(\frac{\ln N}{\ln 2}\right)$ layers. 
We have explained why the system-size nature of the conserved operators of the top layers
is necessary to describe the possible long-ranged spin-glass order of the excited eigenstates and the possible critical points between different FMBL phases. 
We have discussed the similarities and the differences with the Strong Disorder RSRG-X method \cite{rsrgx,rsrgx_moore,vasseur_rsrgx,yang_rsrgx,rsrgx_bifurcation} that constructs the whole set of the $2^N$ eigenstates via a binary tree of $N$ layers.
The block RG construction of the whole spectrum
 has been described for the random quantum Ising chain,
first with short-ranged couplings and then with long-ranged couplings. 
In the Spin-Glass models, we have obtained that the 
excited eigenstates are exactly like ground states in another disorder realization, so that they can be either in the paramagnetic phase, in the spin-glass phase or critical.

\end{document}